\documentclass[10pt]{iopart}
\usepackage{iopams,braket}
\usepackage{graphicx,overpic,floatrow}
\usepackage[caption=false]{subfig}
\begin{document}

\title{Optomechanical cooling by STIRAP-assisted energy transfer: an alternative route towards the mechanical ground state}

\author{Bijita Sarma}
\address{Quantum Systems Unit, Okinawa Institute of Science and Technology Graduate University, Okinawa 904-0495, Japan}
\author{Thomas Busch}
\address{Quantum Systems Unit, Okinawa Institute of Science and Technology Graduate University, Okinawa 904-0495, Japan}
\author{Jason Twamley}
\address{Centre for Engineered Quantum Systems, Department of Physics and Astronomy,
Macquarie University, Sydney, New South Wales 2109, Australia, and}
\address{Quantum Machines Unit, Okinawa Institute of Science and Technology Graduate University, Okinawa 904-0495, Japan}
 
\vspace{10pt}
\begin{indented}
\item[] \today
\end{indented}

\begin{abstract}
Standard optomechanical cooling methods ideally require weak coupling and cavity damping rates which enable the motional sidebands to be well resolved. If the coupling is too large then sideband-resolved cooling is unstable or the rotating wave approximation can become invalid. In this work we describe a protocol to cool a mechanical resonator coupled to a driven optical mode in an optomechanical cavity, which is also coupled to an optical mode in another auxiliary optical cavity, and both the cavities are frequency-modulated. We show that by modulating the amplitude of the drive as well, one can execute a type of STIRAP transfer of occupation from the mechanical mode to the lossy auxiliary optical mode which results in cooling of the mechanical mode. We show how this protocol can outperform normal optomechanical sideband cooling in various regimes such as the strong coupling and the unresolved sideband limit.
\end{abstract}

\vspace{2pc}

\section{Introduction}
Mesoscopic mechanical resonators
have recently garnered extensive theoretical and
experimental research interest due to their potential
uses in quantum information processing and quantum
state engineering \cite{marquardt2009trend,aspelmeyer2014cavity,stannigel2012optomechanical}. In the
field of cavity optomechanics, nanomechanical resonators
have been studied to generate entanglement between
optical and mechanical modes, to facilitate state transfer
between optical and microwave fields, etc., among other various applications. 
However, optomechanical resonators are always in contact with a thermal bath, which hampers the observation of many quantum effects and requires their cooling to the ground state. For this, conventional cavity cooling makes use of optomechanically enhanced damping due to radiation pressure coupling, where the norm is to drive an optomechanical cavity at the red sideband so that the cooling rate can be increased in comparison to the heating rate. In order to resolve the Stokes and anti-Stokes sidebands the cavity decay rate has typically to be much smaller than the mechanical frequency, $\kappa\ll \omega_b$. In this {\em resolved sideband regime} a variety of  optomechanical cooling schemes exist, including ones based on cavity backaction cooling \cite{schliesser2006radiation,gigan2006self,dantan2008self,peterson2016laser}, dissipative optomechanical
coupling \cite{weiss2013quantum,yan2013dissipative}, feedback cooling \cite{mancini1998optomechanical,cohadon1999cooling,kleckner2006sub,corbitt2007optical,poggio2007feedback, genes2008ground, wilson2015measurement},
quadratic coupling \cite{nunnenkamp2010cooling,deng2012performance}, sideband cooling \cite{teufel2011sideband,karuza2012optomechanical},
transient cooling \cite{liao2011cooling,machnes2012pulsed}, cooling based on
the quantum interference effect \cite{xia2009ground,wang2011ultraefficient,genes2008ground},
and others. A few proposals for cooling in the unresolved-sideband regime have been developed as well, based on modulation of the cavity damping rate \cite{elste2009quantum}, using resonant intracavity optical gain \cite{genes2009micromechanical}, optomechanically induced transparency \cite{ojanen2014ground}, feedback cooling \cite{rossi2018measurement}, squeezed light \cite{clark2017sideband,asjad2016suppression}, or by changing photon statistics via parametric interaction \cite{huang2009enhancement}.

Most of the experiments on cavity optomechanical cooling have focused on sideband cooling in the weak coupling regime, which offers the potential to obtain mechanical ground state in the resolved sideband condition.
Nevertheless, the strong optomechanical coupling regime is of interest because it is essential for coherent-quantum control of mechanical resonators, where such resonators can be used for quantum state transfer in optomechanical systems \cite{verhagen2012quantum, groblacher2009observation,palomaki2013coherent,dobrindt2008parametric,akram2010single}, and also for application as quantum transducers for wavelength conversion where it connects to both optical and microwave electromechanical components \cite{o2010quantum,taylor2011laser}. These mechanical control protocols mostly work in precooled optomechanical systems because any quantum fluctuations due to the large thermal bath occupations can deteriorate such state preparations. Hence, it is natural to seek ways to achieve cooling in such strong optomechanical coupling regimes.

\begin{figure}[tb]\label{fig0}
	\centering
	\setlength{\unitlength}{1cm}
	\includegraphics[width=\columnwidth]{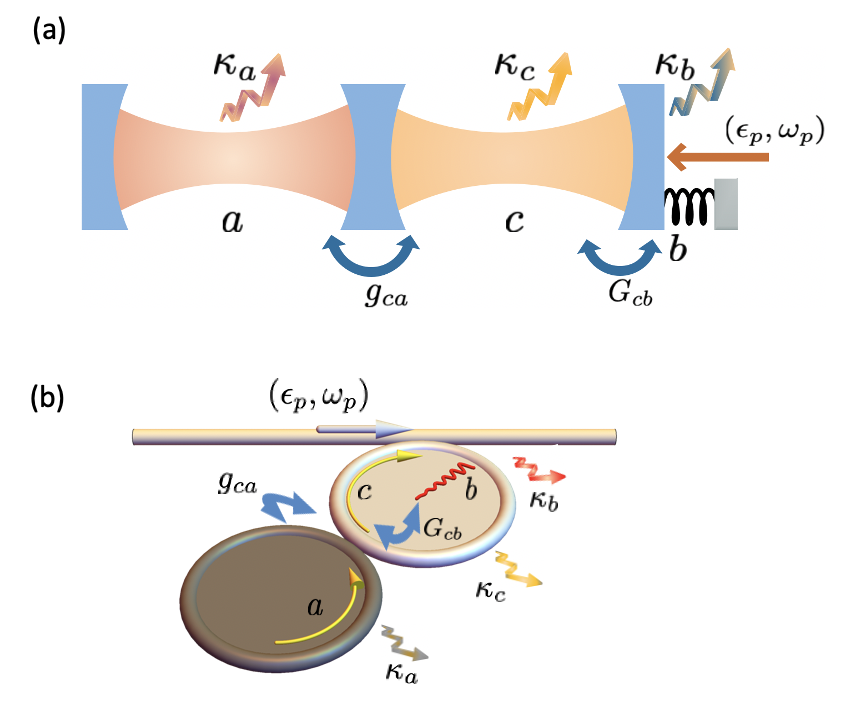}
	\caption{Schematic cooling setups in (a) and (b) show a primary optical cavity with mode $c$, and an auxiliary optomechanical cavity with optical mode $a$, that are coupled at a rate $g_{ca}$. Both the optical modes are frequency-modulated. The primary cavity mode is optomechanically coupled to a mechanical mode $b$, at rate $G_{cb}$, whose strength is modulated via an optical drive on the primary cavity. By modulating the amplitude, $\epsilon_p$, of the optical drive, and the frequencies, $\omega_a$ and $\omega_c$ of the two optical modes, occupation can be transferred from the mechanical mode to the auxiliary cavity mode, thereby cooling the mechanics.}
\end{figure}

In this work we propose a novel method to cool an optomechanical mirror based on adiabatic transfer of phonons into photons. Our model consists of an optomechanical system where a frequency-modulated primary optical cavity, is coupled to a mechanical resonator, and also to a frequency-modulated auxiliary optical cavity. We show that by driving the primary optical cavity with an amplitude-modulated optical field, one can transfer phonons from the mechanical resonator to the auxiliary cavity mode without populating the primary cavity. We use a technique similar to Stimulated Raman Adiabatic Passage (STIRAP) to drain the phononic excitations to the auxiliary lossy optical cavity and by repeatedly iterating the pulses sequence, we show that it is possible to cool the mechanical mode down to the ground state. The advantage of our method is that it operates over a much broader range of conditions than what can be accommodated using standard sideband cooling methods, including strong coupling and the unresolved sideband limit.

\section{Model}
We consider a system consisting of a primary optomechanical cavity coupled to an auxiliary optical cavity, as shown in Fig.~1(a) and (b). There is a cavity-cavity coupling between the primary cavity (with mode $c$) and the auxiliary cavity (with mode $a$), with fixed coupling rate $g_{ca}$. The primary cavity mode is also optomechanically coupled to a mechanical mode ($b$), with single-photon coupling rate $g_{cb}$.
The Hamiltonian of the complete system is given by (in units of $\hbar = 1$)
\begin{eqnarray}
\nonumber
H_0 &=&\omega _{a}a^{\dagger }a + \omega
_{b}b^{\dagger }b + \omega _{c}c^{\dagger }c +g_{cb}c^{\dagger }c (b+b^{\dagger }) + g_{ca}(a^{\dagger }c+c^{\dagger }a)
\label{eq01} \\
&&+i(\varepsilon _{p}c^{\dagger }e^{-i\omega_p t}-\varepsilon_{p}^{\ast }c e^{i\omega_p t}),  
\end{eqnarray}%
where $a(a^{\dagger })$, $b(b^{\dagger })$ and $c(c^{\dagger })$ are the annihilation (creation) operators of the auxiliary cavity mode, mechanical mode and primary cavity mode with resonance frequencies $\omega_a$, $\omega_b$, and $\omega_c$  respectively. The resonance frequencies of the two cavity modes, $\omega_a$ and $\omega_c$, are time-modulated, which can be achieved using an electro-optic modulator \cite{kobayashi1972high, liu2004high, miao2012microelectromechanically, dutt2018experimental}.
The last term describes the external driving of the primary cavity mode, where, $\varepsilon_{p}$ is the amplitude, which will be time-modulated, and $\omega_p$ is the frequency of the drive.
The Hamiltonian of the system in a doubly-rotating frame, under the transformation, $R =\exp \big [ i \omega_p\, (a^\dagger a  + c^\dagger c)\, t \big ]$, with $H=RH_0R^\dagger + i \frac{\partial R}{\partial t}R^\dagger$, is given by
\begin{eqnarray}
\nonumber
H &=&\Delta _{a}a^{\dagger }a + \omega
_{b}b^{\dagger }b + \Delta _{c}c^{\dagger }c +g_{cb}c^{\dagger }c(b+b^{\dagger }) + g_{ca}(a^{\dagger }c + c^{\dagger }a) 
\label{eq02} \\
&&+ i(\varepsilon _{p}c^{\dagger }-\varepsilon_{p}^{\ast }c),  
\end{eqnarray}%
where, $\Delta _{a}(t)=\omega _{a} (t)- \omega _{p}$ and $\Delta _{c}(t)=\omega _{c}(t) - \omega _{p}$, are the cavity detunings. The dynamical evolution of the system operators can be described by the Langevin equations 
\begin{eqnarray}
\dot{a} &=&(-i\Delta _{a}-\kappa _{a})a-ig_{ca}c+\sqrt{2\kappa _{a}}a_{\rm in},
\nonumber \\
\dot{b} &=&(-i\omega _{b}-\kappa _{b})b-ig_{cb}c^{\dagger }c+\sqrt{2\kappa
	_{b}}b_{\rm in},  \\
\dot{c} &=&(-i\Delta _{c}-\kappa _{c})c-ig_{ca}a-ig_{cb}c(b+b^{\dagger })
\label{eq2} +\varepsilon _{p}+\sqrt{2\kappa _{c}}c_{\rm in},  \nonumber
\end{eqnarray}
where $\kappa _{a}$, $\kappa _{b}$ and $\kappa_{c}$ are the losses of
the auxiliary cavity mode, the mechanical mode and the primary cavity mode, respectively. The $a_{\rm in}$, $b_{\rm in}$ and $c_{\rm in}$ are the noise operators
with zero mean values and correlation functions given by $\langle A_{\rm in}(t)A_{\rm in}^{\dagger }(t^{\prime })\rangle =(\bar{n}_{A}+1)\delta (t-t^{\prime })$,  $\langle A_{\rm in}^{\dagger }(t)A_{\rm in}(t^{\prime })\rangle =\bar{n}_{A}\delta (t-t^{\prime })$, and where $\bar{n}_{A}=(e^{\hbar \omega_{A}/k_{B}T_{\rm bath}}-1)^{-1}$ with $A = \{a, b, c\}$, are the mean thermal occupations of the modes. Here $T_{\rm bath}$ is the
common bath temperature and $k_{B}$ is the Boltzmann constant. 
Following the standard linearization procedure for external driving \cite{wang2012using, tian2012adiabatic}, each Heisenberg operator is expressed as a sum of its steady-state mean value and the quantum fluctuation, i.e., $a=\alpha +a_1, b=\beta +b_1$ and $c=\eta +c_1$, where $\alpha$, $\beta$, $\eta$ are the classical mean field values of the modes and $a_1$, $b_1$, $c_1$ are the corresponding quantum fluctuation operators. By separating the classical mean fields and the quantum fluctuations, the classical and quantum Langevin equations can be written as
\begin{eqnarray}
(-i\Delta _{a}-\kappa _{a})\alpha -ig_{ca} \eta&=0,  \nonumber \\
(-i\omega _{b}-\kappa _{b})\beta-ig_{cb}|\eta|^2&=0,  
\\
(-i\tilde{\Delta}_{c}-\kappa _{c})\eta-ig_{ca}\alpha
+\varepsilon_{p} &=0,\nonumber 
\end{eqnarray}
and
\begin{eqnarray}
\dot{a_1} &=&(-i\Delta _{a}-\kappa _{a})a_1-ig_{ca} c_1+\sqrt{2\kappa _{a}}%
a_{in},  \nonumber \\
\dot{b_1} &=&(-i\omega _{b}-\kappa _{b})b_1-ig_{cb}(\eta ^{\ast }c_1+\eta
c^{\dagger }_1)-ig_{cb}c^{\dagger }_1c_1  +\sqrt{2\kappa_{b}}b_{in},  
\\
\dot{c_1} &=&(-i\tilde{\Delta}_{c}-\kappa _{c})c_1-ig_{ca}a_1-ig_{cb}\eta (b_1+b^{\dagger }_1)
-ig_{cb}c_1(b_1+b^{\dagger }_1) +\sqrt{2\kappa _{c}}c_{in}, \nonumber 
\end{eqnarray}%
where $\tilde{\Delta}_{c}=\Delta _{c} + g_{cb}(\beta +\beta ^{\ast })$ with $\beta
=-ig_{cb}\left\vert \eta \right\vert ^{2}/(i\omega _{b}+\kappa _{b})$. For the parameters we consider here, $g_{cb}(\beta +\beta ^{\ast})\ll \Delta _{c}$. Therefore, it can be safely approximated that $\tilde{\Delta}_{c}\approx \Delta _{c}$. The mean field amplitude of the primary cavity mode, $\eta $ is given by
\begin{equation}
\eta =\frac{\varepsilon _{p}(i\Delta _{a}+\kappa _{a})}{g_{ca}^{2}+(i%
	\Delta_{c}+\kappa _{c})(i\Delta _{a}+\kappa _{a})}.  \label{eq021}
\end{equation}%
In the quantum Langevin equations, the product of the fluctuation terms, $ig_{cb}c^{\dagger }_1c_1$ and $ig_{cb}c_1(b_1+b^{\dagger }_1)$, can be considered to be very small in comparison to the other terms, and hence been neglected. Thus, the linearized Hamiltonian of the system is obtained as
\begin{eqnarray}
H_{\rm lin} &=&\Delta _{a}a^{\dagger }_1 a_1+{\Delta}_{c}c^{\dagger }_1c_1+\omega
_{b}b^{\dagger }_1b_1 +G_{cb}(c_1+c^{\dagger }_1)(b_1+b^{\dagger }_1)\nonumber 
\\
&& +g_{ca}(c^{\dagger }_1a_1+c_1a^{\dagger }_1),
\end{eqnarray}%
where $G_{cb}=\eta g_{cb}$ is
the coherent-driving-enhanced linearized optomechanical coupling strength. 
Since $\eta $ is proportional to the amplitude of the driving field, $\varepsilon _{p}$, one can modulate $G_{cb}$ via the external optical drive on $c$. However, it is to be noted that the cavity-cavity coupling $g_{ca}$ cannot be modulated using such a technique and in what follows we will assume that we can time-modulate $G_{cb}$, while $g_{ca}$ is constant in time.

\begin{figure}[!] \label{fig:STIRAP_coh}
	\begin{center}
		\setlength{\unitlength}{1cm}
		\begin{picture}(8.5,6.5)
		\put(-.1,2.0){\includegraphics[width=.48\columnwidth]{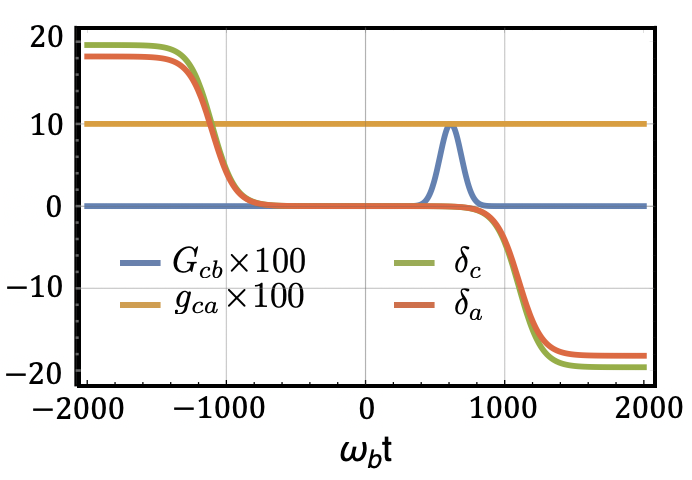}}
		\put(5.3,5.58){(a)}
		\put(6.4,2.0){\includegraphics[width=.48\columnwidth]{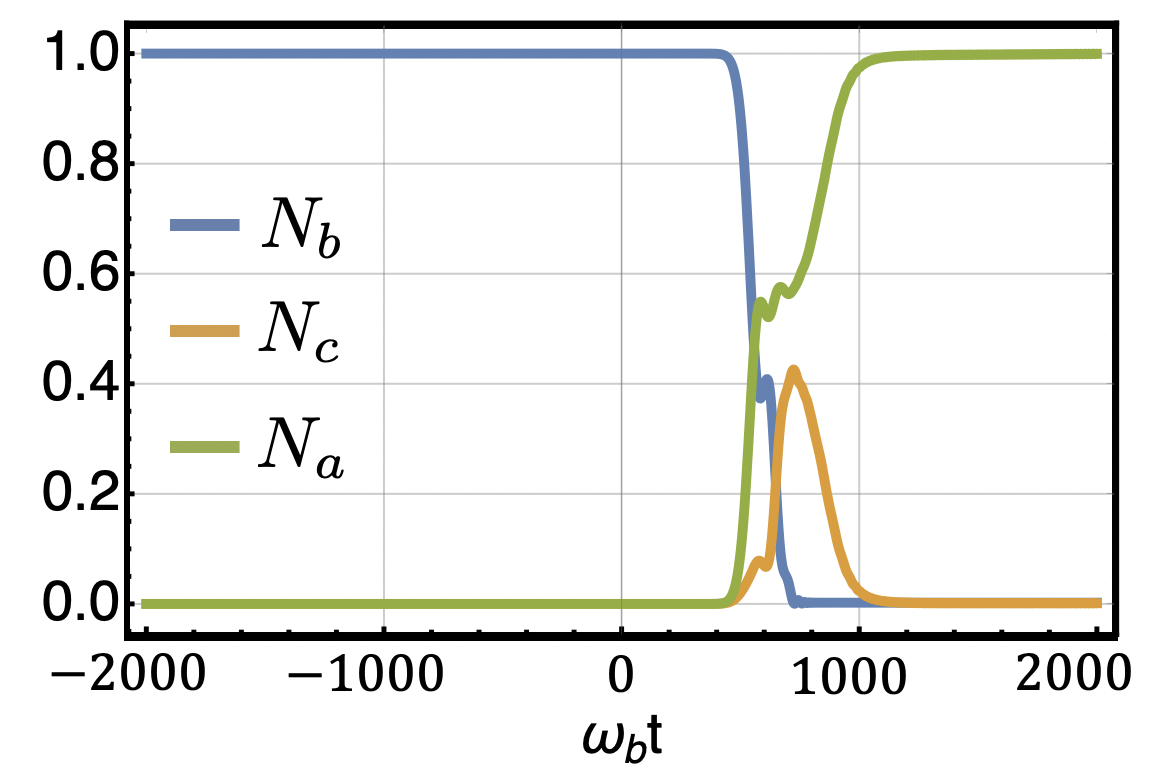}}
		\put(11.8,5.58){(b)}
		\end{picture}
	\end{center}
	\vspace{-2.2 cm}
	\caption{(a) Modulation of the coupling and detuning pulses for the case when $\Omega_0/\omega_b=0.1$. Here $\delta_a,\;\delta_c$ are the cavity detunings and $G_{cb},\;g_{ca}$ are the coupling amplitudes in units of $\omega_b$. (b) Unitary time evolution of the modal populations using the pulses shown in (a), $(N_b, N_c, N_a)$ are the mechanical, primary and auxiliary cavity mode occupations, where initially $(N_b, N_c, N_a)=(1,0,0)$.
		The pulse parameters used are shown in Table \ref{Table1}. 
	}
\end{figure}
\noindent
Transforming the Hamiltonian now to an interaction picture with $U= \exp\left[-i\omega_b (a^\dagger_1 a_1 + c^\dagger_1 c_1 + b^\dagger_1 b_1)t\right]$, yields  $H = U H_{\rm lin} U^\dagger$, where
\begin{eqnarray}
H &=&\delta _{a}(t)a^{\dagger }_1a_1+\delta_{c}(t)c^{\dagger }_1c_1 +G_{cb}(t)\bigg (c^{\dagger }_1b_1+c_1 b^{\dagger }_1   \nonumber \\
&& + e^{-2i\omega_b t} c_1 b_1 + e^{2i\omega_b t} c^\dagger_1 b^\dagger_1 \bigg )  
+g_{ca}(c^{\dagger }_1a_1+c_1a^{\dagger }_1).   \label{RWA1}
\end{eqnarray}%
Here $\delta_a(t) = \Delta_a(t) - \omega_b$ and $\delta_c(t) = {\Delta}_c(t) - \omega_b$ are  time-dependent detunings. One can see that the detunings can be varied by tuning the frequency-modulations of the two cavities, while the optomechanical coupling is varied by tuning the primary cavity drive amplitude. Using these time-dependent modulations we now seek to apply a STIRAP-like protocol to effectively transfer the phonon population to the auxiliary cavity mode. We also note that in most of our analysis below we will {\em not} make the rotating wave approximation (RWA) in the optomechanical coupling term, and the counter rotating terms in the Hamiltonian play an important role particularly when $|G_{cb}|/\omega_b\not \ll 1$.

\section{Population transfer protocol }
In conventional three-level atomic systems, population can be transferred using a Stimulated Raman Adiabatic Passage (STIRAP) protocol, via a so-called `counter-intuitive' modulation of the coupling strengths, to achieve high fidelity transfer between the intial and the final states.
However, this conventional STIRAP method cannot be straightforwardly applied to our system, as the cavity-cavity coupling, $g_{ca}$ cannot be modulated in a time-dependent manner \cite{wang2012using, tian2012adiabatic, di2015population}. 
In what follows we therefore use an alternate method 
which allows 
population transfer from the mechanical mode to the auxiliary cavity mode by modulating the detunings instead, and show how it allows us to cool the mechanical resonator to the ground state. 

\begin{figure}[!] \label{fig:cooling}
	\begin{center}
		\setlength{\unitlength}{1cm}
		\begin{picture}(8.5,8.5)
		\put(-1.55,0.0){\includegraphics[width=\columnwidth]{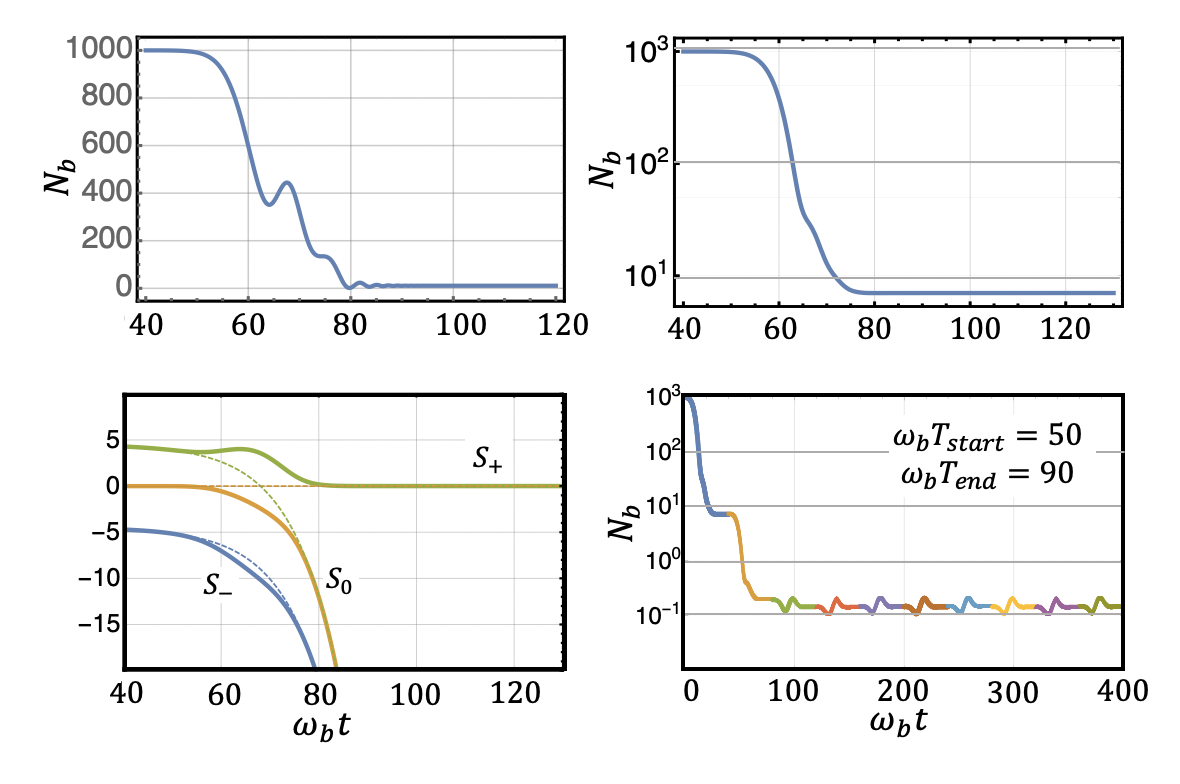}}
		\put(4.0,7.8){(a)}
		\put(10.3,7.7){(b)}
		\put(4.0,1.5){(c)}
		\put(10.3,1.5){(d)}
		\end{picture}
	\end{center}
	\vspace{-0.5 cm}
	\caption{Phonon cooling with and without coupling to thermal baths: 
		(a) Evolution of phonon occupancy in the mechanical resonator, $N_b(t)$, while applying the complete transfer pulses (\ref{eq:pulse})-(\ref{deltas}), with $(N_b, N_a, N_c)=(10^3, 0, 0)$ initially. This is obtained by solving the master equation without considering any damping; (b) Evolution of $N_b(t)$ after including damping in the system, with $(\kappa_c, \kappa_a)/\omega_b = (0.5, 2.0)$, and $Q_b = 10^7$, solved using the master equation approach with initial values $(N_b, N_a, N_c)=(10^3, 0, 0)$; (c) Solid (dashed) lines show the eigenvalues of the time-dependent Hamiltonian with (without) the pump coupling applied, showing the creation of a gap between $S_+$ and $S_0$. (d) $N_b(t)$, when using a truncated pulse ($\omega_bT_{start}=0.5$ to $\omega_bT_{end}=0.9$), which encloses the gap and repeating it 10 times, indicated by different colours used. Parameters for all simulations shown are in Table \ref{Table1}, and we note that here we have considered
		$\Omega_0/\omega_b=0.9$, i.e.~we are in the strong coupling regime where the RWA is no longer valid.}
\end{figure}
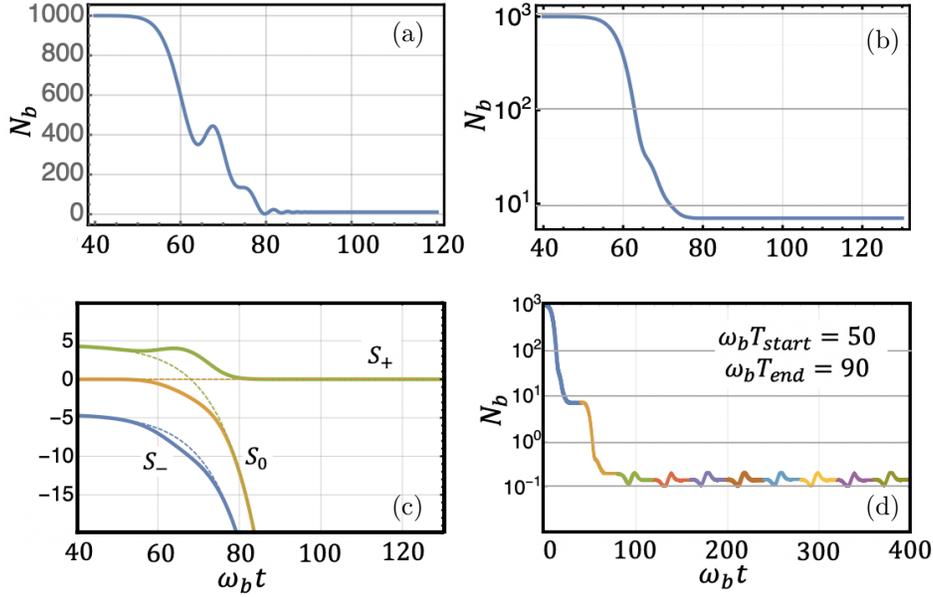
\renewcommand{\arraystretch}{1.3}
\begin{table}[h]
	\caption{\label{Table1} Values of the parameters used in Figs.~2, 3 and 4. The thermal occupations $\bar{n}_a,\,\bar{n}_c$, of the optical modes are taken to vanish, while $\bar{n}_b$ is set equal to the mode's initial occupation, e.g. $\bar{n}_b=10^3$, in Figs.~3 and 4.}
	\small
	\begin{tabular}{|c|c|c|c|c|c|c|c|}
		\hline
		Figure & $\Omega_0/\omega_b$ & $\omega_b\tau_{ch}$ & $\kappa_\delta$ & $\omega_b\tau$ & $h_\delta$  & $\omega_bT$ &  $\omega_bt_c$ \\ \hline
		2 & 0.1 & 164.99 & 14.05 & 1101.69 & 13.94 &  108.76 & 612.26\\ 
		3 & 0.9 & 18.33 & 14.05 &  122.41 & 13.94 & 12.08 	   & 68.03\\ 
		4(a) & 0.3 & 54.99 & 14.05 & 367.23 & 13.94 & 36.25 & 204.08\\
		4(b) & 0.5 & 32.99 & 14.05 & 220.34 & 13.94 & 21.75 & 122.45\\
		4(c) & 0.6 & 27.49 & 14.05 & 183.62 & 13.94 & 18.13 & 102.04\\
		4(d) & 0.9 & 18.33 & 14.05 & 122.41 & 13.94 & 12.08 & 68.03\\
		4(e) & 1.2 & 13.74 & 14.05 & 91.81 & 13.94 & 9.06 & 51.02\\
		4(f) &  0.2 &  82.49 & 14.05 & 550.85 & 13.94 & 54.38 & 306.13\\
		\hline
	\end{tabular}
\end{table}
For this we write the static optical cavity-cavity coupling, which is traditionally known as the `Stokes' coupling, as $g_{ca}\equiv\Omega_s/2=\Omega_0/2$ and set the time dependent optomechanical coupling $G_{cb}(t) \equiv \Omega_p/2$, known as the `Pump coupling', to be the Gaussian 
\begin{equation}
\label{eq:pulse}
\Omega_p(t)= \Omega_0 \mathrm{e}^{-\left(\frac{t-t_c}{T}\right)^2},
\end{equation}
centered at  time $t_c$, with width $T$, and amplitude $\Omega_0$. 
We also apply detunings chosen as
\begin{eqnarray}
\label{eq:detunings}
\nonumber
\delta_c(t) &= \kappa_{\delta} \delta_s(t)\; ,\\
\delta_a(t) &= (\kappa_{\delta} -1)\delta_s(t)\;,
\end{eqnarray}
where
\begin{eqnarray}
\delta_s(t) &=-h_{\delta} \frac{\Omega_0}{2} \left[\tanh\left(\frac{t-\tau}{\tau_{ch}}\right) + \tanh\left(\frac{t+\tau}{\tau_{ch}}\right)\right]\;,\label{deltas} 
\end{eqnarray}
and we will seek values of the parameters ($\kappa_\delta,\; h_\delta,\, \tau,\, \tau_{ch})$, to obtain the best cooling of the $b$-mode for a given strength of driving $\Omega_0/\omega_b$.  The pulse shapes are shown in Fig.~2(a). Here, the parameter $\Omega_s$ is equivalent to a Stokes pulse if one considers an analogous three-level atomic system for normal STIRAP. However, our choice of pulse shape can be better understood by looking at the instantaneous eigenvalues of the system.
In the  rotating wave 
approximation the Hamiltonian  (\ref{RWA1}), can be expressed as 
\begin{eqnarray}
H=
\left[ {\begin{array}{ccc}
	0 								& 	\Omega_p (t)/2							& 0	\\
	\Omega_p^* (t)/2								& 	\delta_c (t)					& \Omega_s/2	\\
	0		& 	\Omega_s/2 	& \delta_a(t) 						\\
	\end{array} } \right]\;\;,
\end{eqnarray}
which has the right form to possess a `dark' eigenstate.
Consider the instantaneous eigenvalues $(\lambda_0, \lambda_1, \lambda_2)$, of this Hamiltonian when the time modulated pulses are applied. If the optomechanical coupling $G_{cb}(t)$  vanishes (i.e.~$\Omega_p(t)=0$), the so-called `Stokes' Hamiltonian is given by
\begin{eqnarray}
H_s=
\left[ {\begin{array}{ccc}
	0 								& 	0							& 0	\\
	0								& 	\delta_c (t)					&  \Omega_0/2	\\
	0		& 	 \Omega_0^*/2	& \delta_a(t) 						\\
	\end{array} } \right]\;\;.
\end{eqnarray}
This Hamiltonian acts only on the 
two cavity subspace, i.e.~it does not involve the mechanical mode, yielding the asymptotic eigenstates $\ket{s_0(t=\pm \infty)}$, and $\ket{s_\pm(t=\pm \infty)}$, where
\begin{eqnarray}
\ket{s_+(-\infty)} &\simeq \ket{N_c}\to\ket{s_+(+\infty)} \simeq \ket{N_a}\;\;,
\\
\ket{s_-(-\infty)} &\simeq \ket{N_a}\to\ket{s_-(+\infty)} \simeq \ket{N_c}\;\;.
\end{eqnarray}
Here $\ket{N_{a}}$ ($\ket{N_{c}}$) are Fock states of the auxiliary (primary) optical cavities and  the corresponding eigenvalues are
\begin{equation}
\label{eq:diabatic-eigenvectors}
S_0 = 0, \quad
S_{\pm} = \frac{\delta_a+\delta_c}{2} \pm \frac{\sqrt{(\delta_a -\delta_c) ^2 + \Omega_0^2}}{2}\;\;.
\end{equation}
The time evolution of the eigenvalues of this Stokes Hamiltonian using the pulses shown in Fig.~2(a), results in the eigenvalues $S_\pm (t)$ crossing the eigenvalue $S_0$ twice at $t\sim \pm t_c$. However, when the Gaussian coupling $\Omega_p$ is applied, it lifts the degeneracy between $S_0$ and $S_+$, resulting in an avoided crossing, which leads to  population transfer. The time evolution of the phonon occupancy in the mechanical resonator, $N_b$ ($\langle b_1^\dagger b_1\rangle$), and photon occupancy in the two cavities, $N_a$ ($\langle a_1^\dagger a_1\rangle$) and $N_c$ ($\langle c_1^\dagger c_1\rangle$), are shown in Fig.~2(b) for the case when initially $(N_b, N_a, N_c)=(1,0,0)$, found by solving the Schr\"odinger equation without considering any  coupling of the system to external baths. 
One can see that the population is transferred with virtually 100\% fidelity from the phonon $b-$mode to the auxiliary cavity $a-$mode.  The population in the primary cavity $c-$mode,  is briefly non-zero and quickly returns to vanishing occupancy, leading to a complete transfer to the auxiliary cavity mode, despite a vast difference in frequencies between the mechanical and optical modes. This method will be extended in the following to study the population dynamics in a realistic open system by coupling each mode to a thermal bath. 

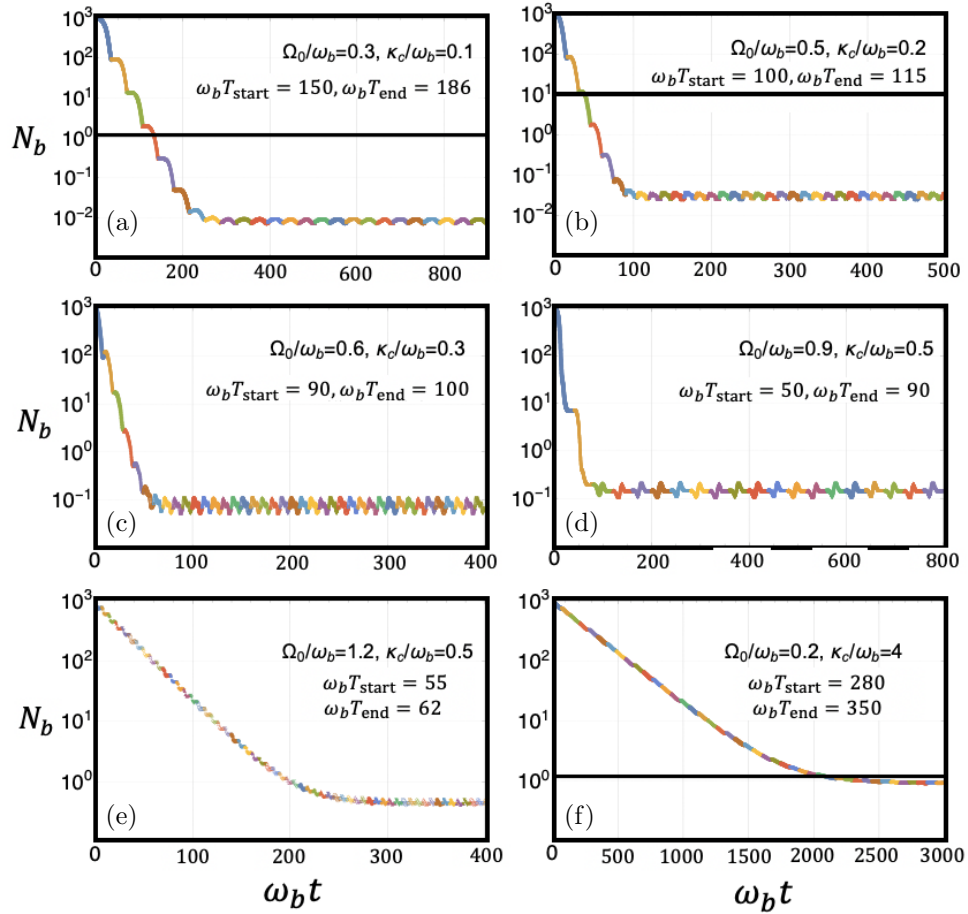
\begin{figure}[tb] \label{fig:comparison}
	\begin{center}
		\setlength{\unitlength}{1cm}
		\begin{picture}(10.5,13)
		\put(-.4,.2){\includegraphics[width=\columnwidth]{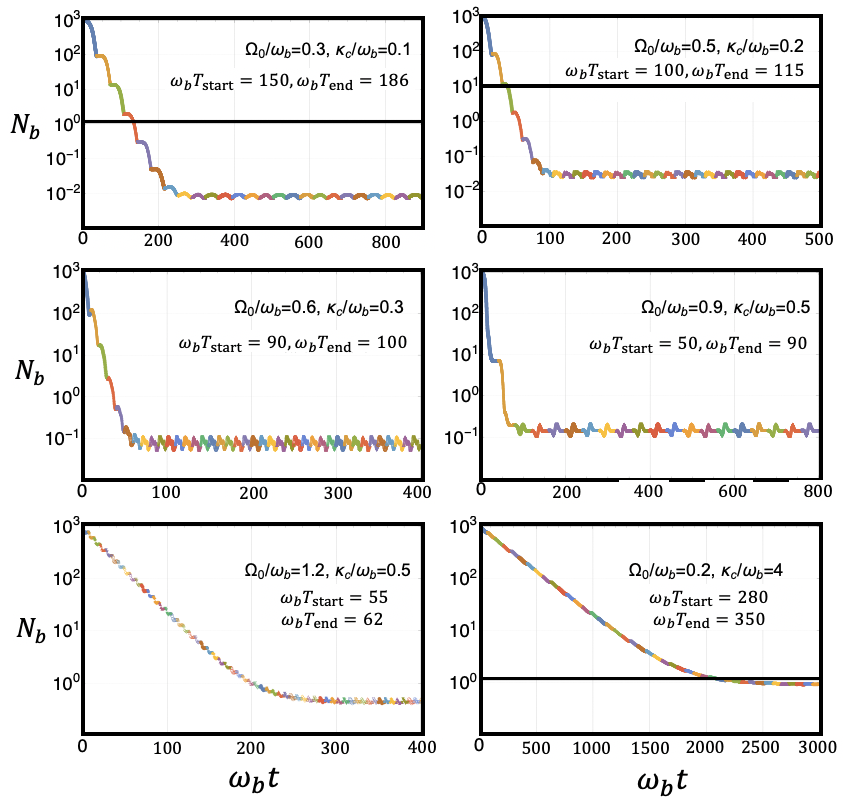}}
		\put(1,9.5){(a)}
		\put(7.1,9.5){(b)}
		\put(1,5.5){(c)}
		\put(7.1,5.5){(d)}
		\put(1,1.6){(e)}
		\put(7.1,1.6){(f)}
		\end{picture}
	\end{center}
	\vspace{0.0 cm}
	\caption{Comparison between optomechanical cooling using iterated STIRAP truncated pulses and standard sideband cooling. Multicoloured curves show phonon occupancy starting from $(N_b, N_a, N_c)=(10^3,0,0)$, using STIRAP cooling for various parameters. The horizontal black solid line (wherever shown), depicts the steady-state phonon number obtained from normal sideband optomechanical cooling while in other cases normal optomechanical cooling is not possible due to instability in the system. 
		The parameters used for each sub-graph are given in Table \ref{Table1}. Here (a)-(e) show cases where the coupling strength $\Omega_0/\omega_b$ gradually increases and also the resolved sideband condition becomes less valid as $\kappa_c/\omega_b$ increases. (f) is the case for moderate coupling strength but is deep in the unresolved sideband regime. Here, $\kappa_a/\omega_b = 2$, $Q_b = 10^7$,  and  $T_{\rm start}$ and $T_{\rm end}$ are the start and end time of each individual truncated sub-pulse. 
	}
\end{figure}
In order to apply our proposed method in a realistic setup one needs to consider  open quantum system dynamics. The phonon number evolution can be studied via covariance methods using the quantum master equation \cite{bienert2015optomechanical, liu2013dynamic}, which for our model is given by
\begin{eqnarray}						
\dot{\rho}= & i\left[\rho, H\right]+
\left\{\kappa_a\left(\bar{n}_{a} +1\right){\cal D}[a_1]+\kappa_a \bar{n}_{a}{\cal D}[a_1^\dagger]\right.\nonumber\\
&+\kappa_c\left(\bar{n}_{c} +1\right){\cal D}[c_1]+\kappa_c \bar{n}_{c}{\cal D}[c_1^\dagger]\nonumber\\
&\left.+\kappa_b\left(\bar{n}_{b} +1\right){\cal D}[b_1]+\kappa_b \bar{n}_{b}{\cal D}[b_1^\dagger]\right\}\rho,
\end{eqnarray}%
where
\begin{eqnarray}
H &=&\delta _{a}(t)a^{\dagger }_1a_1+\delta_{c}(t)c^{\dagger }_1c_1 +G_{cb}(t)(c^{\dagger }_1b_1+c_1 b^{\dagger }_1 + e^{-2i\omega_b t}  \nonumber \\
&&  c_1 b_1+e^{2i\omega_b t} c^\dagger_1 b^\dagger_1 )  
+g_{ca}(c^{\dagger }_1a_1+c_1a^{\dagger }_1),   
\end{eqnarray}%
and ${\cal D}[A]\rho\equiv A\rho A^\dagger-1/2\,\{A^\dagger A,\rho\}$.
We use the covariance approach to find the time evolution of the mean 
phonon number $\langle b^{\dagger }
_1 b_1 \rangle (t)$. For this, we solve a linear system of differential equations
\begin{equation}
{\partial }_t\left\langle {\hat{o}}_i{\hat{o}}_j\right\rangle =Tr\left(\dot{\rho }{\hat{o}}_i{\hat{o}}_j\right)=\sum_{m,n}{{\mu}_{m,n}\left\langle {\hat{o}}_m{\hat{o}}_n\right\rangle }, 
\end{equation} 
where ${\hat{o}}_i$, ${\hat{o}}_j$, ${\hat{o}}_m$, ${\hat{o}}_n$ are one of the 								
operators: ${a}^{\dagger }_1$, $c^{\dagger }_1$, ${b}^{\dagger }_1$, $a_1$, $c_1$ and $b_1$; and ${\mu }_{m,n}$ 		
are the corresponding coefficients. The ordinary differential equations for the time evolution of the second-order moments can be obtained from the master equation as follows,
\begin{eqnarray}
{\partial }_t\langle {a^{\dagger }_1 a_1}\rangle &= & i g_{ca}(\langle {c^{\dagger }_1 a_1}\rangle - \langle {a^{\dagger }_1 c_1}\rangle )  
- \kappa_a (\bar{n}_a + 1) \langle {a^{\dagger }_1 a_1}\rangle \\ \nonumber
&&+ \kappa_a \bar{n}_a (1+ \langle a^\dagger_1 a_1 \rangle ),  \nonumber \\
{\partial }_t\left\langle {c^{\dagger }_1 c_1}\right\rangle &=& i g_{ca} (\langle a^\dagger_1 c_1 \rangle - \langle c^\dagger_1 a_1\rangle) + i G_{cb} (\langle b^\dagger_1 c_1 \rangle - \langle c^\dagger_1 b_1\rangle) + iG_{cb} e^{-2i \omega_b t} \langle c_1b_1 \rangle    \nonumber \\
&& - iG_{cb} e^{2i \omega_b t} \langle c^\dagger_1 b^\dagger_1 \rangle - \kappa_c (\bar{n}_c + 1) \langle {c^{\dagger }_1 c_1}\rangle + \kappa_c \bar{n}_c (1+ \langle c^\dagger_1 c_1 \rangle ),  \nonumber \\
{\partial }_t\langle {b^{\dagger }_1 b_1}\rangle &= & -i G_{cb} (\langle b^\dagger_1 c_1\rangle - \langle c^\dagger_1 b_1\rangle) + iG_{cb} e^{-2i \omega_b t} \langle c_1 b_1 \rangle - iG_{cb} e^{2i \omega_b t} \langle c^\dagger_1 b^\dagger_1 \rangle 
\nonumber \\ 
&&  - \kappa_b (\bar{n}_b + 1) \langle {b^{\dagger }_1 b_1}\rangle  + \kappa_b \bar{n}_b (1+ \langle b^\dagger_1 b_1 \rangle ),  \nonumber \\
{\partial }_t \langle {a^{\dagger }_1 c_1}\rangle &=& i \delta_a \langle  a^\dagger_1 c_1 \rangle -i \delta_c \langle c_1 a^\dagger_1\rangle + i g_{ca} (\langle c^\dagger_1 c_1\rangle  - \langle a^\dagger_1 a_1 \rangle ) - i G_{cb} \langle b_1 a^\dagger_1 \rangle 
\nonumber \\
&&-i G_{cb} e^{2i \omega_b t} \langle a^\dagger_1 b^\dagger_1 \rangle  -(\kappa_a/2) (\bar{n}_a +1 ) \langle a^\dagger_1 c_1 \rangle + (\kappa_a/2) \bar{n}_a \langle c_1 a^\dagger_1 \rangle   \nonumber \\
&&-  (\kappa_c/2) (\bar{n}_c +1 ) \langle a^\dagger_1 c_1 \rangle + (\kappa_c/2) \bar{n}_c \langle c_1 a^\dagger_1 \rangle,
\nonumber \\
{\partial }_t \langle {a^{\dagger }_1 b_1}\rangle &= & i \delta_a \langle a^\dagger_1 b_1 \rangle + i g_{ca} \langle c^\dagger_1 b_1 \rangle - i G_{cb} \langle c_1 a^\dagger_1 \rangle  -i G_{cb} e^{2i \omega_b t} \langle a^\dagger_1 c^\dagger_1 \rangle   \nonumber \\
&& -(\kappa_a/2) (\bar{n}_a +1 ) \langle a^\dagger_1 b_1 \rangle   
+ (\kappa_a/2) \bar{n}_a \langle a^\dagger_1 b_1\rangle- (\kappa_b/2) (\bar{n}_b +1 ) \langle a^\dagger_1 b_1 \rangle \nonumber \\
&&+ (\kappa_b/2) \bar{n}_b \langle a^\dagger_1 b_1 \rangle , 
\nonumber \\
{\partial }_t \langle {c^{\dagger }_1 b_1} \rangle &= & i  \delta_c \langle c^\dagger_1 b_1\rangle + i g_{ca} \langle a^\dagger_1 b_1 \rangle 
+ i G_{cb} (\langle b^\dagger_1 b_1 \rangle- \langle c^\dagger_1 c_1 \rangle) + i G_{cb} e^{-2i \omega_b t} \langle b_1 b_1 \rangle 
\nonumber \\
&&- i G_{cb} e^{2i \omega_b t} \langle c^\dagger_1 c^\dagger_1 \rangle   -(\kappa_b/2) (\bar{n}_b +1 ) \langle c^\dagger_1 b_1 \rangle + (\kappa_b/2) \bar{n}_b \langle  c^\dagger_1 b_1 \rangle  
 \nonumber \\
 &&-  (\kappa_c/2) (\bar{n}_c +1 ) \langle c^\dagger_1 b_1 \rangle 
 + (\kappa_c/2) \bar{n}_c \langle  c^\dagger_1 b_1 \rangle,
\nonumber \\
{\partial }_t \langle {b_1 b_1 } \rangle &= &  - 2 i G_{cb} \langle b_1 c_1 \rangle - 2 i G_{cb} e^{2i \omega_b t} \langle c^\dagger_1 b_1 \rangle + \kappa_b (\bar{n}_b+1)  \langle b_1 b_1 \rangle \nonumber \\	
&&+ \kappa_b \bar{n}_b  \langle b_1 b_1  \rangle , 
\nonumber \\	
{\partial }_t\left\langle {c_1 b_1}\right\rangle &= &  - i \delta_c \langle b_1 c_1 \rangle-i G_{cb} (\langle c_1 c_1 \rangle + \langle b_1 b_1 \rangle) -i g_{ca} \langle a_1 b_1    \rangle 
\nonumber \\
&&- (\kappa_b/2) (\bar{n}_b+1) \langle b_1 c_1 \rangle + (\kappa_b/2) \bar{n}_b \langle b_1 c_1  \rangle  -iG_{cb} e^{2i \omega_b t} (\langle c^\dagger_1 c_1 \rangle  \nonumber \\
&&+ \langle b^\dagger_1 b_1 \rangle +1)- (\kappa_c/2) (\bar{n}_c+1) \langle b_1 c_1 \rangle + (\kappa_c/2) \bar{n}_c \langle b_1 c_1  \rangle ,
\nonumber \\
{\partial }_t \langle {a^\dagger_1 c^\dagger_1} \rangle &= & i  (\delta_a + \delta_c) \langle  a^\dagger_1 c^\dagger_1 \rangle + g_{ca} (\langle  c^\dagger_1 c^\dagger_1 \rangle + \langle  a^\dagger_1 a^\dagger_1 \rangle ) + G_{cb} \langle  a^\dagger_1 b^\dagger_1 \rangle 
 \nonumber \\
&&+ G_{cb} e^{-2i \omega_b t}  \langle  a^\dagger_1 b_1\rangle  - (\kappa_a/2) (\bar{n}_a+1) \langle a^\dagger_1 c^\dagger_1 \rangle  \nonumber \\
&&+ (\kappa_a/2) \bar{n}_a \langle a^\dagger_1 c^\dagger_1   \rangle - (\kappa_c/2) (\bar{n}_c+1) \langle a^\dagger_1 c^\dagger_1 \rangle + (\kappa_c/2) \bar{n}_c \langle a^\dagger_1 c^\dagger_1  \rangle , 
\nonumber \\
{\partial }_t \langle {a^\dagger_1 b^\dagger_1} \rangle &= & i \delta_a \langle  a^\dagger_1 b^\dagger_1 \rangle + i g_{ca}  \langle  c^\dagger_1 b^\dagger_1 \rangle  + i G_{cb}  \langle  a^\dagger_1 c^\dagger_1 \rangle + i G_{cb} e^{-2i \omega_b t}  \langle  a^\dagger_1 c_1 \rangle 
\nonumber \\
&&- (\kappa_a/2) (\bar{n}_a+1) \langle a^\dagger_1 b^\dagger_1 \rangle  + (\kappa_a/2) \bar{n}_a \langle a^\dagger_1 b^\dagger_1 \rangle \nonumber \\
&&- (\kappa_b/2) (\bar{n}_b+1) \langle a^\dagger_1 b^\dagger_1 \rangle + (\kappa_b/2) \bar{n}_b \langle a^\dagger_1 b^\dagger_1  \rangle , 
\nonumber \\
{\partial }_t \langle {c^\dagger_1 c^\dagger_1 } \rangle &= &   2i \delta_c \langle c^\dagger_1 c^\dagger_1  \rangle  + 2 i g_{ca} \langle a^\dagger_1 c^\dagger_1  \rangle + 2 i G_{cb} \langle b^\dagger_1 c^\dagger_1   \rangle + 2 i G_{cb} e^{-2i \omega_b t} \langle c^\dagger_1 b_1  \rangle \nonumber \\
&&+ \kappa_c (\bar{n}_c+1)  \langle c^\dagger_1 c^\dagger_1 \rangle+ \kappa_c \bar{n}_c  \langle c^\dagger_1 c^\dagger_1 \rangle , 
\nonumber \\
{\partial }_t \langle {a^\dagger_1 a^\dagger_1 } \rangle &= & 2 i \delta_a \langle a^\dagger_1 a^\dagger_1  \rangle +2 i g_{ca} \langle a^\dagger_1 c^\dagger_1  \rangle 
+ \kappa_a (\bar{n}_a+1) \langle a^\dagger_1 a^\dagger_1 \rangle+ \kappa_a \bar{n}_a \langle a^\dagger_1 a^\dagger_1 \rangle . 
\end{eqnarray}
Solving these equations, one can determine the 	
mean values of all the time-dependent second-order moments: $\langle a^{\dagger }_1 a_1 \rangle $, $\langle c^{\dagger }_1 c_1\rangle $, $\langle b^{\dagger }_1 b_1\rangle $, $\langle a^{\dagger }_1 c_1\rangle $, $\langle a^\dagger_1 b_1\rangle $, $\langle c^\dagger_1 b_1\rangle$, $\langle c_1 b_1 \rangle $, $\langle a^{\dagger }_1 b^{\dagger }_1 \rangle $,  $\langle c^{\dagger }_1 a^{\dagger }_1 \rangle $, $\langle b_1^2 \rangle $, $\langle c^{\dagger }_1 c^{\dagger }_1 \rangle $ and $\langle a^{\dagger }_1 a^{\dagger }_1 \rangle$. 
In the following, we will consider an initial state of the system where only the $b-$mode is occupied, e.g.~$\langle b^{\dagger }_1 b_1\rangle (t=0)$ is nonzero. We will consider that at $t=0$, all the other second-order moments vanish. In particular the initial thermal occupations of the optical cavities at room temperatures is assumed to be vanishingly small.  

\begin{table}[h]
	\caption{\label{jlab1}Comparison of steady-state phonon number calculated for normal cooling $N_{min}^{NC}$, and the minimal phonon number obtained using the iterated STIRAP-cooling method $N_{min}^{SC}$, for a variety of parameters. Here, $\omega_b T_{\rm start}$ and $\omega_bT_{\rm end}$ are the start and end time of each pulse which have been found to achieve the optimal cooling in each case. We observe that STIRAP cooling succeeds in all cases and in some cases reaches lower final phonon occupations. }

	\footnotesize
	\begin{tabular}{|c|c|c|c|c|c|c|c|}
		\hline
		$G/\omega_b$&$\kappa_c/\omega_b$&$\kappa_a/\omega_b$&$Q_b$&$\omega_bT_{\rm start}$&$\omega_b T_{\rm end}$&$N_{min}^{NC}$ &$N_{min}^{SC}$\\
		\hline
		0.02&0.05&0.01&$10^5$&3000&3180&0.51&2.11\\ 
		0.02&0.05&2&$10^5$&3000&3180&0.51&1.98\\
		0.02&0.05&0.01&$10^7$&3000&3180&0.005&0.021\\
		0.02&0.05&2&$10^7$&3000&3180&0.005&0.020\\	
		0.1&0.1&0.01&$10^5$&500&600&0.13&0.52\\	
		0.1&0.1&2&$10^5$&500&600&0.13&0.39\\	
		0.1&0.1&0.01&$10^7$&500&600&0.0070&0.0077\\		
		0.1&0.1&2&$10^7$&500&600&0.0070&0.0058\\
		0.3&0.1&0.01&$10^5$&150&186&0.173&0.442\\
		0.3&0.1&2&$10^5$&150&186&0.173&0.219\\
		0.3&0.1&0.01&$10^7$&150&186&0.071&0.016\\
		0.3&0.1&2&$10^7$&150&186&0.071&0.008\\		
		0.5&0.2&0.01&$10^5$&100&115&12.679&0.184\\		
		0.5&0.2&2&$10^5$&100&115&12.679&0.114\\		
		0.5&0.2&0.01&$10^7$&100&115&12.628&0.046\\		
		0.5&0.2&2&$10^7$&100&115&12.628&0.030\\		
		0.6&0.3&0.01&$10^5$&90&100&unstable&0.179\\
		0.6&0.3&2&$10^5$&90&100&unstable&0.136\\
		0.6&0.3&0.01&$10^7$&90&100&unstable&0.097\\
		0.6&0.3&2&$10^7$&90&100&unstable&0.080\\		
		0.9&0.5&0.01&$10^5$&50&90&unstable&0.300\\		
		0.9&0.5&2&$10^5$&50&90&unstable&0.266\\		
		0.9&0.5&0.01&$10^7$&50&90&unstable&0.161\\	
		0.9&0.5&2&$10^7$&50&90&unstable&0.149\\
		1.2&0.5&0.01&$10^5$&55&62&unstable&1.574\\
		1.2&0.5&2&$10^5$&55&62&unstable&0.717\\
		1.2&0.5&0.01&$10^7$&55&62&unstable&0.441\\
		1.2&0.5&2&$10^7$&55&62&unstable&0.451\\	
		1.5&0.5&0.01&$10^5$&44&51&unstable&1.574\\		
		1.5&0.5&2&$10^5$&44&51&unstable&1.44\\		
		1.5&0.5&0.01&$10^7$&44&51&unstable&1.143\\	
		1.5&0.5&2&$10^7$&44&51&unstable&1.03\\		
		0.2&4&0.01&$10^5$&280&350&1.273&3.512\\
		0.2&4&2&$10^5$&280&350&1.273&3.46\\
		0.2&4&0.01&$10^7$&280&350&1.023&0.953\\
		0.2&4&2&$10^7$&280&350&1.023&0.940\\		
		0.5&10&0.01&$10^5$&100&150&6.480&10.09\\		
		0.5&10&2&$10^5$&100&150&6.480&9.91\\	
		0.5&10&0.01&$10^7$&100&150&6.381&5.37\\		
		0.5&10&2&$10^7$&100&150&6.381&5.277\\
		\hline
	\end{tabular}
\end{table}

Using this approach, we plot the unitary evolution of the phonon occupation, $N_b$,  for a system in the strong coupling regime with
$\Omega_0/\omega_b=0.9$ in Fig.~3(a), without considering any damping in the system.  Setting initially $(N_b, N_a, N_c)=(10^3, 0, 0)$, we see that one can achieve nearly perfect transfer out of the $b$-mode. In order to consider a realistic system, we incorporate damping in the system with $\kappa_c/\omega_b = 0.5,\;\kappa_a/\omega_b = 2$, and $Q_b$ $(\omega_b/\kappa_b) = 10^7$, and  initially $N_b=10^3$, and we observe the evolution shown in Fig.~3(b). Although the phonon occupation reduces significantly it does not reach the ground state.  

In Fig.~3(c) we plot the eigenspectrum and note that when the pump pulse is applied, a gap opens up between the eigenstates $S_0$ and $S_+$. As the system evolves adiabatically from the initial state to the final state, it stays on the same eigenstate because of this avoided crossing \cite{shore2017picturing,vitanov2017stimulated,bergmann2019roadmap}, and it is this gap that permits the transfer. In STIRAP, the system evolution along the eigenstate is perfect if the evolution is adiabatic and infinitely slow. Such infinitely slow evolution will permit perfect transfer despite the gap to another eigenstate becoming very small (as long as the gap is nonzero), anytime during the adiabatic evolution. However, for faster evolution, non-adiabatic evolution might occur which will degrade the transfer \cite{shore2017picturing}. If the gap can be widened then such non-adiabatic processes are greatly reduced improving the transfer. And as can be seen from Eq.~(16), the eigenvalue $S_+$ increases as $\Omega_0$ is increased; thereby opening up the gap between $S_0$ and $S_+$, allowing faster transfers, but there are physical limits on how large $\Omega_0/\omega_b$ can be \cite{verhagen2012quantum, groblacher2009observation,palomaki2013coherent,dobrindt2008parametric,akram2010single}.
Since most of the transfer occurs during this gap we consider in the following a truncated portion of the full pulse chosen from the behaviour in the interval $t\in \{T_{start},T_{end}\}$, which closely matches the temporal location of this gap.   Iterating this truncated sub-pulse a number of times, as shown in Fig.~3(d), allows us to minimise the time for heating and thereby efficiently cool the mechanical mode to its ground state.  In the following, we will apply this method over a range of system parameters, and we will also discuss the advantages over standard optomechanical cooling.
\vspace{0cm}\\[2em]
\section{Comparison to standard optomechanical cooling} In standard optomechanical sideband cooling, a  quantum cooling limit exists which is characterised by when the system finally attains a stationary state, i.e.~$\mathrm{d}\langle o_{i}o_{j}\rangle/\mathrm{d}t=0$. When working on the red side-band  $(\Delta_{c}=\omega_{b})$, and when the cooperativity parameter $C\equiv4|G_{cb}|^{2}/(\kappa_b\kappa_c)\gg1$, the steady-state final mean phonon number is given by \cite{Bai2018},
\begin{eqnarray}\label{e010}
\langle b^{\dagger}_1 b_1\rangle_{\mathrm{lim}}& \simeq & \frac{4|G_{cb}|^{2}+\kappa_c^{2}}{4|G_{cb}|^{2}(\kappa_c+\kappa_b)}\kappa_b \bar{n}_b \nonumber \\
&& +\frac{(4\omega_{b}^{2}-\kappa_c^{2})(8|G_{cb}|^{2}+\kappa_c^{2})+2\kappa_c^{4}}{16\omega_{b}^{2}(4\omega_{b}^{2}
	+\kappa_c^{2}-16|G_{cb}|^{2})},
\end{eqnarray}
where the first term describes the cooling limit in the presence of a motional thermal environment, while the second term describes the  cooling limit achieved in the case where the motional bath is the vacuum. This latter is non-zero as the cooling process itself has competing cooling and heating rates. The stability condition is given by the Routh-Hurwitz criterion, $|G_{cb}|^2<\omega_b^2/4+\kappa_c^2/16$ \cite{Bai2018}.

In Fig.~4, we compare the reduction of the phonon occupancy achieved via iterated STIRAP pulses and standard sideband cooling in different regimes. For the parameters shown in Figs.~4(c), (d) and (e), the conditions for stability in normal sideband optomechanical cooling are violated and that cooling method fails.
However, our method succeeds and one can almost reach the ground state in most cases. For the parameters shown in Figs.~4(a), (b) and (f), normal sideband cooling works, however, it is evident that our method returns better cooling in these regimes. An elaborate comparison is presented in Table 2 for a range of parameters from where the regimes where our method succeeds over normal sideband cooling can be easily identified. It can be seen that in the unresolved sideband regime, i.e.~$\kappa_c\gg \omega_b$,  cooling with our STIRAP pulses can be improved with higher $\kappa_a/\omega_b$.

\section{Conclusions}
Cooling of mechanical resonators remains a crucial goal in the engineering of quantum motional states of matter. Using a detuning-assisted STIRAP scheme we have shown that a cooling method exists which effectively transfers the quanta from the mechanical oscillator to an optical oscillator in a one-way fashion and operates over a broad range of parameters. Just as normal STIRAP transfer is quite robust to pulse/parameter imperfections we expect our scheme should also exhibit similar robustness.  

\section*{Acknowledgements}
We acknowledge support from the ARC Centre of Excellence for Engineered Quantum Systems CE170100009 and the Okinawa Institute of Science and Technology Graduate University.

\section*{References}

\providecommand{\newblock}{}

\end{document}